# Carbon-Rich Carbon Nitride Monolayers with Dirac Cones: Dumbbell C$_4$N

Linyang Li[a,*], Xiangru Kong[b,*], Ortwin Leenaerts[a,*], Xin Chen[c], Biplab Sanyal[c], François M. Peeters[a,*]

[a] *Department of Physics, University of Antwerp, Groenenborgerlaan 171, B-2020 Antwerp, Belgium*

[b] *International Center for Quantum Materials, Peking University, 100871 Beijing, China*

[c] *Department of Physics and Astronomy, Uppsala University, SE-75120 Uppsala, Sweden*

**Abstract:** Two-dimensional (2D) carbon nitride materials play an important role in energy-harvesting, energy-storage and environmental applications. Recently, a new carbon nitride, 2D polyaniline (C$_3$N) was proposed [PNAS 113 (2016) 7414-7419]. Based on the structure model of this C$_3$N monolayer, we propose two new carbon nitride monolayers, named dumbbell (DB) C$_4$N-I and C$_4$N-II. Using first-principles calculations, we systematically study the structure, stability, and band structure of these two materials. In contrast to other carbon nitride monolayers, the orbital hybridization of the C/N atoms in the DB C$_4$N monolayers is $sp^3$. Remarkably, the band structures of the two DB C$_4$N monolayers have a Dirac cone at the K point and their Fermi velocities (2.6/2.4×10$^5$ m/s) are comparable to that of graphene. This makes them promising materials for applications in high-speed electronic devices. Using a tight-binding model, we explain the origin of the Dirac cone.

*Corresponding authors. E-mail addresses:
linyang.li@uantwerpen.be (Linyang Li)
kongxru@pku.edu.cn (Xingru Kong)
ortwin.leenaerts@uantwerpen.be (Ortwin Leenaerts)
francois.peeters@uantwerpen.be (François M. Peeters)

## 1. Introduction

Monolayer graphene was first realized in 2004 [1] and since then the group IV elemental monolayers have played a crucial role in the field of two-dimensional (2D) materials. In experiments, silicene, germanene, and stanene were all synthesized on different substrates [2-10], such as heterostructures of silicene(germanene)/$MoS_2$ [6,8] and stanene/$Bi_2Se_3$ [10], which preserve the hexagonal honeycomb structure of the isolated monolayers. Without spin-orbit coupling (SOC), silicene, germanene, and stanene all have a zero electronic band gap with two bands crossing linearly at the Fermi level and their extremely large Fermi velocity ($v_F$) makes them ideal materials for high-speed electronic devices [11,12]. Taking SOC into account, they all show a nontrivial band gap at the K point which can reach up to 73.5 meV (stanene), making them promising to realize quantum spin Hall (QSH) effect at room-temperature [12]. Besides the buckled hexagonal honeycomb structure, another stable structure model of Si/Ge/Sn, known as the dumbbell (DB) structure, was proposed in previous studies [13-15]. Different to the band structure of stanene, the DB structure of Sn shows a band inversion at the Γ point due to SOC and was predicted to be a 2D topological insulator [13]. By functionalization, Sn/Ge DB structures show nontrivial band gaps that can reach up to 235 meV [16-19] while the Si DB structure can also become a 2D topological insulator under external strain [20]. In contrast to the extended literature on DB structures of Si/Ge/Sn, studies on the DB structure of C have been scarce up to now [21].

Besides the group IV elemental monolayers and their allotropes, carbon nitride materials are another important set of 2D materials. Many 2D carbon nitride materials can be applied in some important physical and chemical processes. Among these, graphitic carbon nitride (g-$C_3N_4$) has been studied for a long time [22] and it can be used in many energy applications, such as hydrogen generation

[23], efficient energy storage [24], and photocatalytic degradation of pollutants [25]. Nanoporous carbon nitride materials, such as the $C_2N$ monolayer [26], can be applied in gas separation [27-29] or water desalination [30]. The carbon nitride materials not only have a huge potential in applications but also exhibit many interesting physical effects, including QSH [31], quantum anomalous Hall (QAH) [32,33], and spin-polarization [34,35] effects. More and more new 2D carbon nitride materials have been produced in experiment. Recently, an interesting new 2D carbon nitride material, called polyaniline ($C_3N$), was synthesized [36]. Its monolayer shows a similar structure as graphene and its band structure is semiconducting with an indirect band gap, in contrast to the Dirac band structure of graphene.

Based on the $C_3N$ monolayer structure and the DB structure of Si/Ge/Sn, we propose two new $C_4N$ monolayers with a DB structure (DB $C_4N$). According to the positions of the raised C/N atoms, two configurations, DB $C_4N$-I and DB $C_4N$-II, can be obtained. Using first-principles calculations, we investigated systematically the structure, energy, stability, and electronic band structure of these two DB $C_4N$ monolayers. Different to other carbon nitride monolayers, all the C and N atoms in the DB $C_4N$ monolayers have $sp^3$ hybridization. Although the two DB $C_4N$ monolayers have a different structure and ground state energy, the phonon spectra provide convincing evidence for their thermal and dynamical stability. Similar to the Dirac cone band structure of graphene, the two DB $C_4N$ monolayers both show a Dirac cone at the K point with a large Fermi velocity. An analysis of the projected electron density of states (PDOS) and the electron wave functions of the Dirac cones shows that four $p_z$ atomic orbitals are responsible for the Dirac cone in these two $C_4N$ monolayers. This is also supported by our tight-binding (TB) model including four $p_z$ atomic orbitals that reproduces the first-principles results quite well.

**2. Calculation Method**

Our first-principles calculations were performed using the Vienna *ab initio* simulation package (VASP) code [37-39], implementing density functional theory (DFT). The electron exchange-correlation functional was treated using the generalized gradient approximation (GGA) in the form proposed by Perdew, Burke, and Ernzerhof (PBE) [40]. The atomic positions and lattice vectors were fully optimized using the conjugate gradient (CG) scheme until the maximum force on each atom was less than 0.01 eV/Å. The energy cutoff of the plane-wave basis was set to 520 eV with an energy precision of $10^{-5}$ eV. The Brillouin zone (BZ) was sampled by using a 17×17×1 Γ-centered Monkhorst-Pack grid. The vacuum space was set to at least 20 Å in all the calculations to minimize artificial interactions between neighboring slabs. The phonon spectra were calculated using a supercell approach within the PHONOPY code [41].

## 3. Results and Discussions

### 3.1 *Geometrical Structure*

The two investigated DB $C_4N$ monolayers are shown in Figure 1(a) and (b). They are based on the $C_3N$ monolayer (Figure 1(c)), which was recently synthesized in experiment [36]. By adsorbing C atoms on all the N atom positions of the $C_3N$ monolayer, we obtain two different DB $C_4N$ monolayers. The two patterns are classified as follows: (1) If the raised N/C atoms are on one side of the monolayer, as shown in Figure 1(a), we call it DB $C_4N$-I; (2) If the raised N/C atoms are on two opposite sides of the monolayer, as shown in Figure 1(b), we call it DB $C_4N$-II. In the following, we will analyze the three monolayers (DB $C_4N$-I, DB $C_4N$-II, and $C_3N$) from the points of view of symmetry, orbital hybridization style, and bond length.

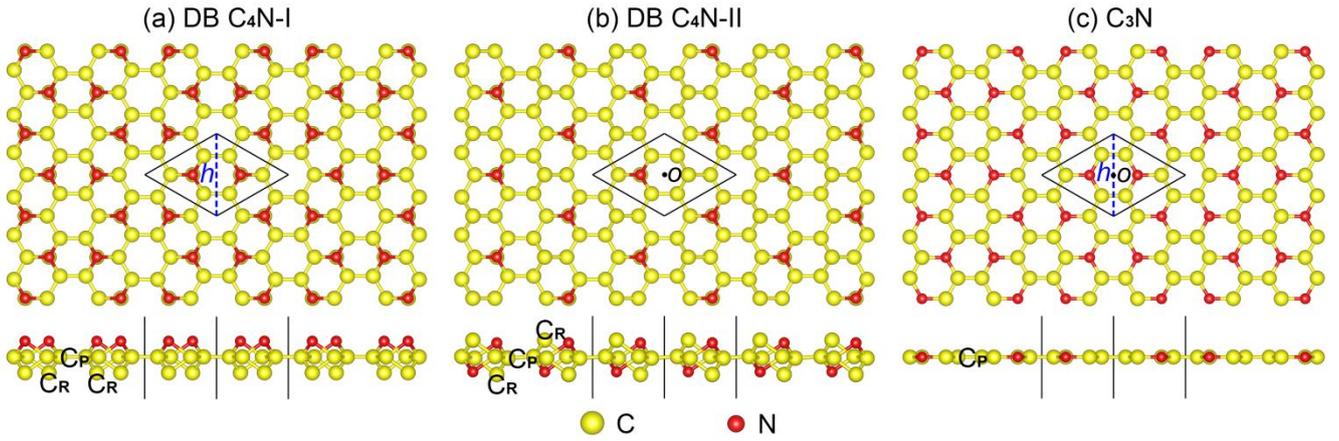

**Figure 1.** Schematic representations (top and side views) of the DB $C_4N$-I (a), DB $C_4N$-II (b) and $C_3N$ (c). The raised C atoms are labeled as $C_R$ and the planar C atoms are labeled as $C_P$. The yellow (red) symbols are the carbon (nitrogen) atoms.

Considering symmetry first, the $C_3N$ monolayer has higher symmetry than the $C_4N$ monolayers. For the $C_3N$ monolayer, the C/N atoms are not only center symmetric about the point *o* (black point, Figure 1(c)), but also mirror symmetric about the plane *h* (blue dashed line, Figure 1(c)) parallel to the *z* axis. However, the C/N atoms of the DB $C_4N$-I are only mirror symmetric about the plane *h* (blue dashed line, Figure 1(a)) parallel to the *z* axis while the C/N atoms of the DB $C_4N$-II are only center symmetric about the point *o* (black point, Figure 1(b)). Next, let us take a look at the hybridization of the atoms. The C/N atoms in the $C_3N$ monolayer have $sp^2$ hybridization, similar to the atoms in graphene or g-$C_3N_4$. On the other hand, all the C/N atoms in the DB $C_4N$ monolayers have $sp^3$ hybridization, similar to the C atoms in diamond.

Corresponding to the change in hybridization, their bond lengths are very different. There are three kinds of atoms in the DB $C_4N$ monolayers, N atoms, raised C ($C_R$) atoms, and planar C ($C_P$) atoms, while there are only N atoms and $C_P$ atoms in the $C_3N$ monolayer. In the following, we will discuss the bonds between the different atoms. Considering the N-$C_P$ bond, one N atom can form a bond with three $C_P$ atoms and the lengths of the three bonds are the same, labeled as *L*. For the DB $C_4$N-I and $C_4$N-II, *L* = 1.558 Å and *L* = 1.545 Å, respectively, which is larger than the *L* = 1.403 Å in $C_3$N. Between the C atoms, only $C_P$-$C_P$ bonds are found in the $C_3$N monolayer and we label them as *M* (*M* = 1.404 Å). Similar bonds are also found in the DB $C_4$N monolayers. The $C_P$ atoms of the DB $C_4$N-I are all in the same *xy* plane. However, there is a little buckling (0.095 Å) along the *z* axis between the $C_P$ atoms in the DB $C_4$N-II due to the inequivalence of the N and $C_R$ atoms. The *M* of the DB $C_4$N-I and $C_4$N-II is 1.491 Å and 1.488 Å, respectively. There is also another C-C bond in the DB $C_4$N monolayers between the $C_R$ atom and the $C_P$ atom, which is labeled as *M'*. The *M'* of the DB $C_4$N-I ($C_4$N-II) is 1.567 Å (1.583 Å). From the above data of bond lengths, we can summarize that all the bond lengths in the two DB $C_4$N monolayers are 1.49~1.58 Å, which is much larger than the 1.40 Å in the $C_3$N monolayer. A similar difference is also found in the C-C bond length of diamond (1.54 Å) [42] and graphene (1.42 Å) [43], which comes from the different hybridization of the C atoms ($sp^3$ and $sp^2$). In most experimental and predicted carbon nitride monolayers, the C and N atoms have $sp^2$/$sp$ hybridization, while the carbon nitride monolayers with $sp^3$-hybridized C/N atoms are rarely studied [44]. Here, we provide a novel structure model for new stable carbon nitride monolayers. Although there is much difference in bond length, the lattice constants of the $C_3$N (4.861 Å) and the DB $C_4$N (4.775 Å ($C_4$N-I) and 4.768 Å ($C_4$N-II)) are almost the same. The optimized geometrical structure data are summarized in Table 1.

Table 1. Structure parameters and formation energies of the DB $C_4$N and $C_3$N monolayers. *a* is the lattice constant. *L*, *M*, and *M'* are the lengths of the N-$C_P$, $C_P$-$C_P$, and $C_P$-$C_R$ bonds, respectively. The unit of $a$, $L$, $M$, and $M'$ is Å. The formation energy $\Delta E$ corresponds to the energy release of the $C_3N$ monolayer adsorbing isolated magnetic C atoms (eV/C atom).

|  | $a$ | $L$(N-$C_P$) | $M$($C_P$-$C_P$) | $M'$($C_P$-$C_R$) | $\Delta E$ |
|---|---|---|---|---|---|
| $C_4N$-I | 4.775 | 1.558 | 1.491 | 1.567 | -3.272 |
| $C_4N$-II | 4.768 | 1.545 | 1.488 | 1.583 | -3.379 |
| $C_3N$ | 4.861 | 1.403 | 1.404 | -- | -- |

*3.2 Energy and Stability*

We calculated the formation energies of the DB $C_4N$ monolayers as the difference in energy between the DB $C_4N$ and the sum of the energies of the $C_3N$ monolayer and isolated magnetic C atoms [45]. The formation energy $\Delta E$ of the DB $C_4N$-I ($C_4N$-II) is -3.272 eV/C atom (-3.379 eV/C atom), which is of the same magnitude as adsorbed C atoms on graphene [46]. The negative formation energy suggests the stability and feasibility of the DB $C_4N$. In experiment, carbon atoms and carbon atomic chains have been observed on graphene using transmission electron microscope (TEM) and TEM studies demonstrated that individual carbon atoms can be adsorbed on graphene surfaces with a stable structure at room-temperature [47-48]. Since 2D $C_3N$ has been realized in experiment, the DB $C_4N$ monolayers are probably also suited for experimental synthesis.

Considering the two DB $C_4N$ monolayers, the energy of the DB $C_4N$-II is lower than that of the DB $C_4N$-I. This can be ascribed to the differences in their structures. The DB $C_4N$-I monolayer exhibits a polarized structure due to the absence of inversion symmetry while the DB $C_4N$-II has no dipole moment. From the views of energy and symmetry, the DB $C_4N$-II is more stable than the DB $C_4N$-I. Although the two DB $C_4N$ monolayers have different ground state energies, their phonon spectra (shown in Figure 2) are both free from imaginary frequency modes, which indicates that they are dynamically stable.

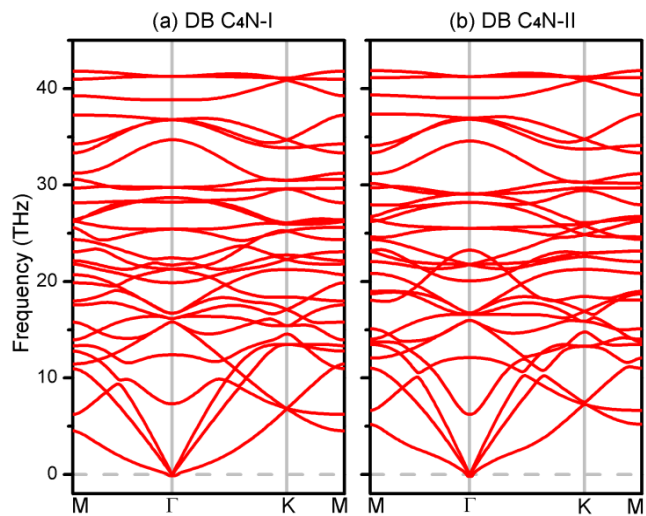

**Figure 2**. Phonon spectra of DB $C_4N$-I (a) and DB $C_4N$-II (b) along the high-symmetry lines in the BZ.

*3.3 Band Structure*

The electronic band structure of the DB $C_4N$-I and $C_4N$-II monolayers is shown in Figure 3. For both cases, a zero band gap at the K point is seen. Two bands cross linearly at the Fermi level and the charge carriers can be characterized by massless Dirac fermions, similar to the band structure of graphene (silicene/germanene/stanene) at the K point [11,12]. By a linear fitting of the first-principles data, we can obtain the Fermi velocity of the DB $C_4N$. For the DB $C_4N$-I ($C_4N$-II), the Fermi velocity is $2.6 \times 10^5$ m/s ($2.4 \times 10^5$ m/s), which is comparable to that of the group IV elemental monolayers ($4.70 \sim 8.46 \times 10^5$ m/s) [12]. The extremely large Fermi velocity makes the DB $C_4N$ monolayers ideal materials for building high-speed electronic devices, such as field effect transistors (FET). It is well known that there are many kinds of Dirac carbon monolayers in experiment and theory, such as graphene [1], α/β/δ/6,6,12-graphyne [49-51], phagraphene [52], etc. However, the 2D carbon nitride materials in experiment, such as g-$C_3N_4$, $C_2N$, and g-$C_6N_6$ monolayers, all show semiconductor behavior with a band gap of 2.73 eV [53], 1.96 eV [26], and 1.53 eV [31], respectively. Dirac cone band structures for carbon nitride materials are rarely predicted and only the g-$C_{14}N_{12}$ and g-$C_{10}N_9$ were proposed as spin-polarized Dirac materials [32,33]. To our knowledge, the two DB $C_4N$ monolayers are the first predicted Dirac carbon nitride materials without spin-polarization. In contrast to the g-$C_6N_6$ monolayer proposed by Wang *et al.* [31], the Dirac cone of the DB $C_4N$ is right at the Fermi level, which is advantageous for experiments and applications.

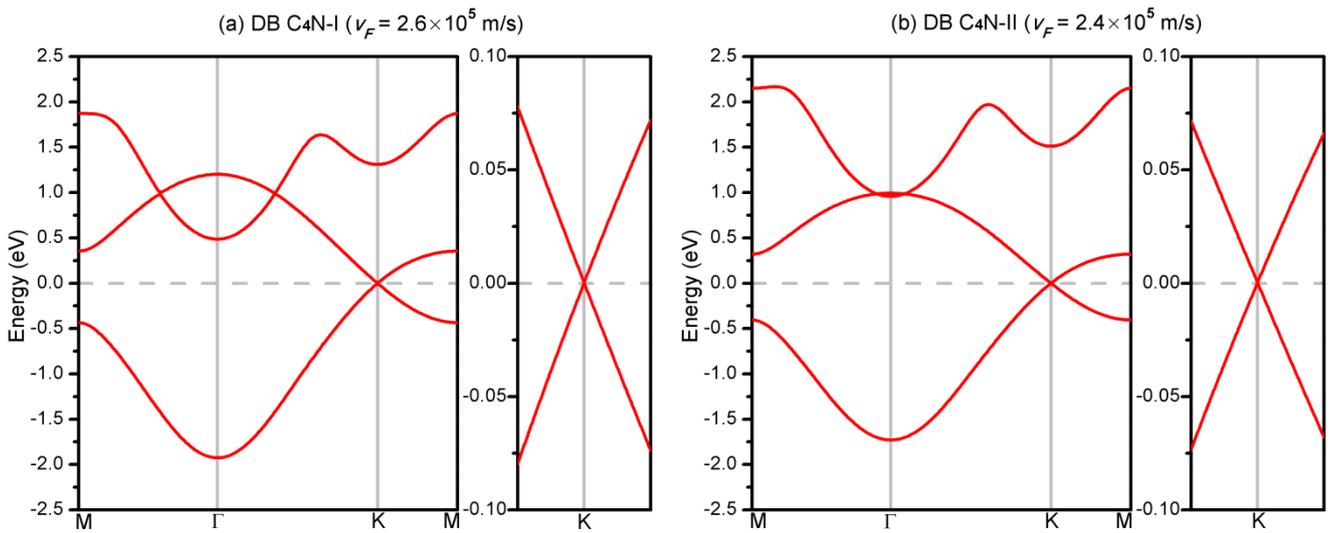

**Figure 3**. Band structure of DB $C_4N$-I (a) and DB $C_4N$-II (b) monolayers from DFT calculations. $v_F$ is the Fermi velocity of the Dirac cone. The energy at the Fermi level was set to zero. An enlarged view of the band lines at the K point near the Fermi level is presented on the right.

To investigate the origin of the Dirac cones, we calculated the PDOS for different atomic orbitals and the Kohn-Sham wave functions at the Dirac points, as shown in Figure 4. From the PDOS analysis of the two DB $C_4N$ monolayers, it is clear that the summed states of the $s$, $p_x$, and $p_y$ atomic orbitals around the Fermi level are far less than the PDOS corresponding to the $p_z$ atomic orbitals. Although the C/N atoms are $sp^3$-hybridized $((sp^3)_z = 1/2s + \sqrt{3}/2p_z)$, the states close to the Fermi level mainly come from the $p_z$ atomic orbitals, which is similar to graphene/graphyne [49-52]. As for phagraphene where only the $p_z$ orbitals of eight out of twenty C atoms in the unit cell are responsible for the formation of the Dirac cone [54], not all ten atoms in the unit cell of the DB $C_4N$ contribute to the Dirac cone. Only the two $C_R$ and the two N atoms play an important role in the formation of the Dirac cone, which is clearly shown in the corresponding Kohn-Sham wave functions at the Dirac point (Figure 4(a) and (b)).

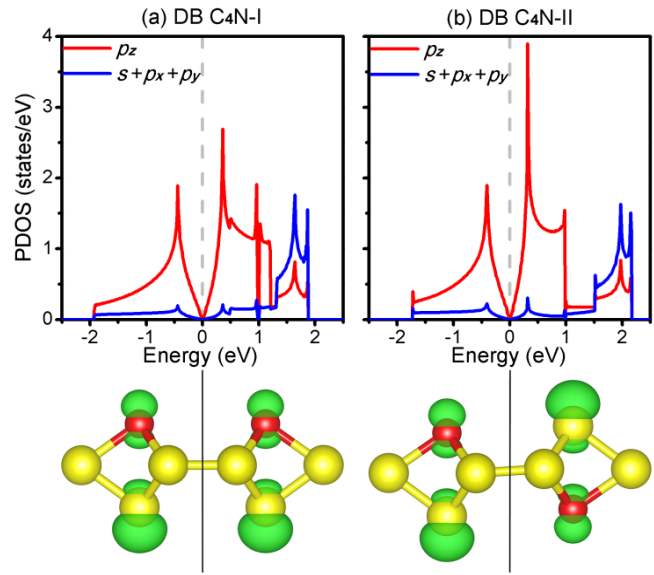

**Figure 4.** The PDOS for the summed states of the $s$, $p_x$, and $p_y$/$p_z$ atomic orbitals of all the atoms in the unit cell and the corresponding Kohn-Sham wave functions at the Dirac points. The isosurface values of the Kohn-Sham wave functions were set to 0.02 Å$^{-3}$.

We also investigated the SOC effect on the band structures of the DB $C_4N$ monolayers. For the DB

C$_4$N monolayers, the SOC effect is very small with the value of the band gap at the K point below 0.25 meV. This will be difficult to be detected in a real experiment because of the possible presence of disorder. We can understand the weak SOC effect from two aspects. First, the SOC effect is related to the element type: the heavier the element, the more pronounced the SOC effect becomes. Structures including heavy elements, such as Bi, can have a large SOC effect [55, 56]. Second, the SOC effect is also related to the atomic orbitals forming the Dirac cone. For the g-C$_6$N$_6$ monolayer, the SOC band gap of the Dirac cone at the K point (below the Fermi level) can reach 5.50 meV [31]. In this case the Dirac cone is formed by the $p_x$ and $p_y$ atomic orbitals of the N atoms and the SOC gap originates directly from the onsite term. According to the PDOS and the corresponding Kohn-Sham wave functions of the Dirac point, the four $p_z$ atomic orbitals of the C$_R$ and N atoms in the DB C$_4$N are responsible for the formation of the Dirac cone, which is the main reason for the weak SOC effect. The band gap values of the DB C$_4$N monolayers are between the 0.008 meV of graphene and the 0.59 meV of δ-graphyne [51], where the Dirac cones are formed by the $p_z$ atomic orbitals of the C atoms.

*3.4 TB model*

Although the two C$_R$ and the two N atoms all contribute to the Dirac cone, it is clear that the Kohn-Sham wave functions of the C$_R$ atoms are much larger than those of the N atoms for both DB C$_4$N monolayers (Figure 4). We can understand the origin of the Dirac cone from a simple TB model by only including the $p_z$ atomic orbitals of the two C$_R$ atoms without the two N atoms. The two DB C$_4$N monolayers can be regarded as to consist of two C$_R$ atoms in one unit cell. Setting $\vec{a}_1$ and $\vec{a}_2$ as the primitive vectors, the positions of the two C$_R$ atoms in the *xy* plane can be expressed as $1/3 \times (\vec{a}_1 + \vec{a}_2)$ and

$2/3 \times (\vec{a}_1 + \vec{a}_2)$, similar to graphene/silicene. The DB $C_4N$-I can be regarded as a structure of graphene because the two $C_R$ atoms are in the same $xy$ plane while the DB $C_4N$-II can be regarded as a structure of silicene due to the buckling along the $z$ axis between the two $C_R$ atoms. The two $C_R$ atoms in the unit cell can be labeled as A and B atoms. The TB Hamiltonian that describes the electronic structure near the Fermi level of such a system can be written as [57,58]

$$H = \begin{pmatrix} \varepsilon & \hbar v_F(k_x - ik_y) \\ \hbar v_F(k_x + ik_y) & -\varepsilon \end{pmatrix},$$

where $k$ is the wave vector relative to the Dirac point and $v_F$ is the Fermi velocity. Similar to graphene/silicene, the difference of the onsite energy between the two $C_R$ atoms is zero because they have the same chemical environment [59,60]. Thus we obtain a linear dispersion relation $E = \pm \hbar v_F |k|$, which is the origin of the Dirac cone.

Although the above simple model leads to a direct understanding of the Dirac cone, the two N atoms still have contributions in the formation of the Dirac cone. We further confirm these results by a TB model including the $p_z$ atomic orbitals of the two $C_R$ and two N atoms in the unit cell [54]. We resorted to maximally localized Wannier functions (MLWFs) [61] using the Wannier90 code [62]. Starting from the bands of the first-principles calculations, we used the four $p_z$ atomic orbitals of the $C_R$ and N atoms. Minimizing the MLWF spread, the band structures obtained using the Wannier90 interpolation method and the DFT calculations are in excellent agreement, as shown in Figure 5.

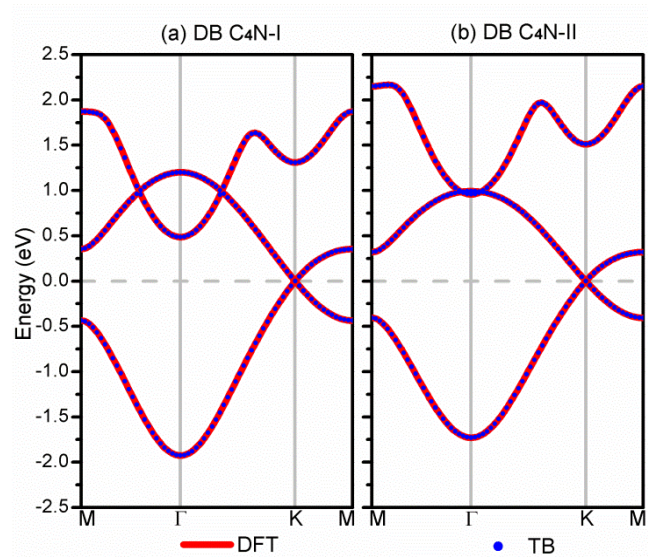

**Figure 5.** Band structure of DB C$_4$N-I (a) and DB C$_4$N-II (b) from DFT and TB calculations. The red lines represent the results of DFT calculations and the blue dotted lines represent the results of TB calculations.

## 4. Conclusion

In summary, using first-principles calculations, we predict the first Dirac carbon nitride monolayers without spin-polarization. In contrast to the existing 2D carbon nitride materials that have been realized in experiments, they have a dumbbell structure and all the C and N atoms are $sp^3$-hybridized. Their formation energies compared to C$_3$N and phonon calculations fully confirm their energetic and dynamical stability. The Dirac cone band structure and large Fermi velocity are comparable to graphene and make the DB C$_4$N monolayers promising materials for high-speed electronic devices. A simple TB model was constructed in order to understand the origin of the Dirac cone which is helpful for further investigations of the massless Dirac fermions. The here proposed DB C$_4$N monolayer structures provide a new way to search for Dirac cone band structures in group IV-V elemental materials.

**Acknowledgements**

This work was supported by the Fonds Wetenschappelijk Onderzoek (FWO-Vl). The computational resources and services used in this work were provided by the VSC (Flemish Supercomputer Center), funded by the Research Foundation - Flanders (FWO) and the Flemish Government – department EWI. X. C. acknowledges financial support from the China Scholarship Council (CSC).